\documentclass[aps,prl,reprint,superscriptaddress,amsmath,amssymb]{revtex4-2}
\usepackage{graphicx}
\usepackage{dcolumn}
\usepackage{bm}
\usepackage{todonotes} 
\usepackage{physics}
\usepackage{siunitx}
\usepackage[colorlinks=true, linkcolor=blue, citecolor=blue, urlcolor=blue]{hyperref}
\usepackage{siunitx}
\usepackage{ulem}

\usepackage{braket}

\usepackage{color}
\usepackage{colordvi}  
\definecolor{Red}{rgb}{1,0,0}

\makeatletter
\def\authornote{\xdef\@thefnmark{$\dagger$}\@footnotetext}
\makeatother

\begin{document}

\title{Coherent Dynamics of the Swing-Up Excitation Technique}

\author{Katarina Boos$^\dagger$}
 \affiliation{Walter Schottky Institut, TUM School of Computation, Information and Technology, and MCQST, Technische Universit\"at M\"unchen, 85748 Garching, Germany}%
\author{Friedrich Sbresny$^\dagger$}
 \affiliation{Walter Schottky Institut, TUM School of Computation, Information and Technology, and MCQST, Technische Universit\"at M\"unchen, 85748 Garching, Germany}%
\author{Sang Kyu Kim$^\dagger$}
 \affiliation{Walter Schottky Institut, TUM School of Computation, Information and Technology, and MCQST, Technische Universit\"at M\"unchen, 85748 Garching, Germany}%
 \authornote{These authors contributed equally.}
\author{Malte Kremser}
 \affiliation{Walter Schottky Institut, TUM School of Natural Sciences, and MCQST, Technische Universit\"at M\"unchen, 85748 Garching, Germany}%
\author{Hubert Riedl}
 \affiliation{Walter Schottky Institut, TUM School of Natural Sciences, and MCQST, Technische Universit\"at M\"unchen, 85748 Garching, Germany}%
\author{Frederik W. Bopp}
 \affiliation{Walter Schottky Institut, TUM School of Natural Sciences, and MCQST, Technische Universit\"at M\"unchen, 85748 Garching, Germany}%
\author{William Rauhaus}
 \affiliation{Walter Schottky Institut, TUM School of Computation, Information and Technology, and MCQST, Technische Universit\"at M\"unchen, 85748 Garching, Germany}%
\author{Bianca Scaparra}
 \affiliation{Walter Schottky Institut, TUM School of Computation, Information and Technology, and MCQST, Technische Universit\"at M\"unchen, 85748 Garching, Germany}%
\author{Klaus D. J\"ons}
 \affiliation{Institute for Photonic Quantum Systems (PhoQS), Center for Optoelectronics and Photonics Paderborn (CeOPP) and Department of Physics, Paderborn University, 33098 Paderborn, Germany}%
\author{Jonathan J. Finley}
 \affiliation{Walter Schottky Institut, TUM School of Natural Sciences, and MCQST, Technische Universit\"at M\"unchen, 85748 Garching, Germany}%
\author{Kai M\"uller}
 \affiliation{Walter Schottky Institut, TUM School of Computation, Information and Technology, and MCQST, Technische Universit\"at M\"unchen, 85748 Garching, Germany}%
\author{Lukas Hanschke}
 \affiliation{Institute for Photonic Quantum Systems (PhoQS), Center for Optoelectronics and Photonics Paderborn (CeOPP) and Department of Physics, Paderborn University, 33098 Paderborn, Germany}%
 
\date{\today}

\begin{abstract}
Developing coherent excitation methods for quantum emitters which enable high brightness, good single-photon purity and high indistinguishability of the emitted photons has been a key challenge in the past years. While many methods have been proposed and explored, they all have specific advantages and disadvantages. In this letter, we investigate the dynamics of the recent swing-up scheme as an excitation method for a two-level system and its performance in single-photon generation. By applying two far red-detuned laser pulses, the two-level system can be prepared in the excited state with near-unity fidelity. We demonstrate the successful functionality and the coherent character of this technique using semiconductor quantum dots. Moreover, we explore the multi-dimensional parameter space of the two laser pulses to study the impact on the excitation fidelity. Finally, we investigate the performance of the scheme as an excitation method for generation of high-quality single photons. We find that the swing-up scheme itself works well and exhibits nearly perfect single-photon purity, while the observed indistinguishability in our sample is limited by the influence of the inevitable high excitation powers on the semiconductor environment of the quantum dot.
\end{abstract}

\pacs{Valid PACS appear here}

\maketitle

Semiconductor quantum dots (QDs) are among the most promising candidates for applications in photonic quantum technologies~\cite{Senellart2017, Gao2012}. They are under close investigation as a material system for optically-active spin qubits for photonic quantum gates~\cite{Sun2016}, remote-entanglement~\cite{delteil2016,stockill2017}, or the generation of highly-entangled photon graph states~\cite{Schwartz2016, istrati2020}. Furthermore, especially as single-photon sources they outperform other solid-state quantum emitters in terms of brightness and quality of the emitted photons~\cite{Gazzano2013, Schweickert2018,Tomm2021}. For the generation of single photons, in recent years many resonant and non-resonant methods for excitation have been established, such as cross-polarized resonant excitation~\cite{He2013, Kuhlmann2015}, phonon-assisted excitation ~\cite{Ardelt2014, Reindl2019, Thomas2021}, excitation via the p-shell~\cite{ware2005, Cogan2021}, dichromatic excitation~\cite{He2019, Koong2021}, or excitation via the biexciton~\cite{Brunner1994, Schweickert2018, Hanschke2018}. Even though each of these methods is promising in its own way, they come along with specific individual disadvantages. For example, resonant excitation requires challenging filtering~\cite{vamivakas2009, flagg2009, muller2007} and results in higher multi-photon emission probability~\cite{fischer2017}, while non-resonant excitation introduces additional timing-jitter which limits the indistinguishability of the emitted photons~\cite{Simon2005, huber2013}. Other limitations of some of the schemes are low efficiency or simply the restriction to a specific level scheme like the presence of a cascaded level structure~\cite{Schweickert2018, Sbresny2022}. Moreover, for applications where the coherence of the system has to be maintained, coherent excitation is crucial.
Recently, Bracht \textit{et al.}~\cite{Bracht2021} proposed a promising technique which combines coherent and off-resonant excitation for a two-level system, whose principle functionality was confirmed experimentally by Karli \textit{et al.}~\cite{Karli2022}. This swing-up excitation method is based on two far red-detuned laser pulses which swing up the population from ground to excited state for a specific relation in their multi-dimensional space - frequency, intensity, pulse duration and timing. This elegant way of exciting the two-level system directly is highly promising as the theoretical model predicts near-unity fidelity and good quality of the emitted photons while enabling frequency filtering and high brightness.

In this Letter, we explore the swing-up excitation method experimentally and theoretically with regard to the multi-dimensional parameter space. We find several resonances and good agreement between our experimental results and numerical simulations. Furthermore, we compare the characteristics of the emitted photons to the ones generated via resonant excitation in terms of lifetime, single-photon purity and indistinguishability.

\begin{figure}
    \centering
    \includegraphics[width=\linewidth]{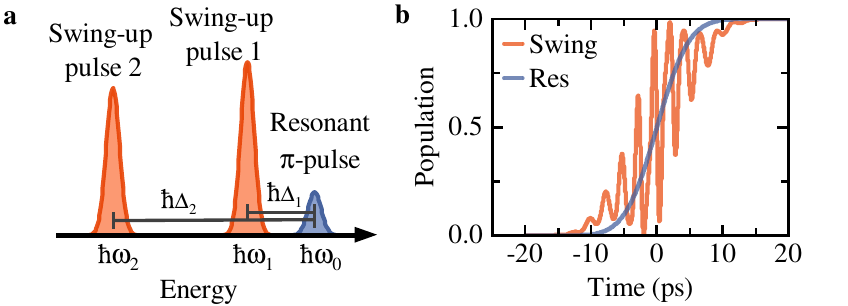}
    \caption{Schematic illustration and simulation of the swing-up excitation. (a) Qualitative energy separation of the far red-detuned pulses for swing-up excitation (orange) and the resonant pulse (blue). (b) Simulated time-resolved state occupation for swing-up (orange) and resonant $\pi$-pulse population inversion (blue) during the excitation process.}
    \label{fig:graph1}
\end{figure}
In our study, the two-level system is realized by the transition of the negatively charged exciton in a QD, with one single electron $e^-$ in the QD as the ground state $\ket{0}$ and the trion $X^-$ as the excited state $\ket{1}$. The transition energy between excited and ground state is given by $\hbar \omega_0$ while the laser energies are given by $\hbar \omega_{i}= \hbar \omega_0 + \hbar \Delta_i$ with $i=1,2$, where $\hbar \Delta_i$ is the detuning from resonance. The system can be easily inverted from $\ket{0}$ to $\ket{1}$ with near-unity efficiency under resonant excitation with a $\pi$-pulse area as schematically illustrated in blue in Fig.~\ref{fig:graph1}b~\cite{stievater2001}. While a single red-detuned laser pulse with $\hbar \Delta < 0$ is incapable of efficiently transferring population, certain combinations of two red-detuned pulses enable near-unity population inversion~\cite{Bracht2021}. For the swing-up scheme, the system Hamiltonian in the rotating frame oscillating at the center frequency of the less detuned laser pulse and after applying the rotating wave approximation is
\begin{equation}
H = - \hbar \Delta_1 \sigma ^\dagger \sigma + \frac{1}{2}\hbar( \Omega^\ast (t) \sigma + \Omega (t) \sigma ^\dagger)
\end{equation}
where the two-color excitation field is described by  
\begin{equation}
\Omega (t) = \Omega_1 (t) + \Omega_2 (t+\tau) e^{-i(\omega_2-\omega_1) t}
\end{equation}
with Gaussian envelopes $\Omega_i (t)$ and time delay $\tau$ between the pulses. The integrated pulse areas are given by $\alpha_i = \int_{-\infty}^{+\infty}\Omega_i(t)\,dt$. The excitation pulses are schematically depicted in Fig.~\ref{fig:graph1}a in orange. Note that the phase between the pulses does not affect the scheme and is thus set to zero~\cite{Bracht2021}. 
By solving the von Neumann equation $\frac{\partial}{\partial t} \rho (t) = - \frac{i}{\hbar} [H, \rho (t)]$ where $\rho$ is the density matrix, we can calculate the population of the excited state~\cite{Bracht2021}. Details on the simulations are discussed in the supplemental material ~\cite{Supplemental}. As shown by the simulation results in Fig.~\ref{fig:graph1}b in orange, the population can be inverted with near-unity fidelity, where the occupation is not smoothly transferred but shows an oscillatory, upswinging behaviour during the presence of the pulse.

For our measurements, we used a single \mbox{InGaAs} QD embedded in a Schottky diode structure. The diode stabilizes the charge environment in the vicinity of the QD and enables deterministic switching between neutral and negatively charged QD ground states~\cite{warburton2000, seidl2005}. Further information on sample and experimental setup can be found in the supplemental material~\cite{Supplemental}.

\begin{figure}
    \centering
    \includegraphics[width=\linewidth]{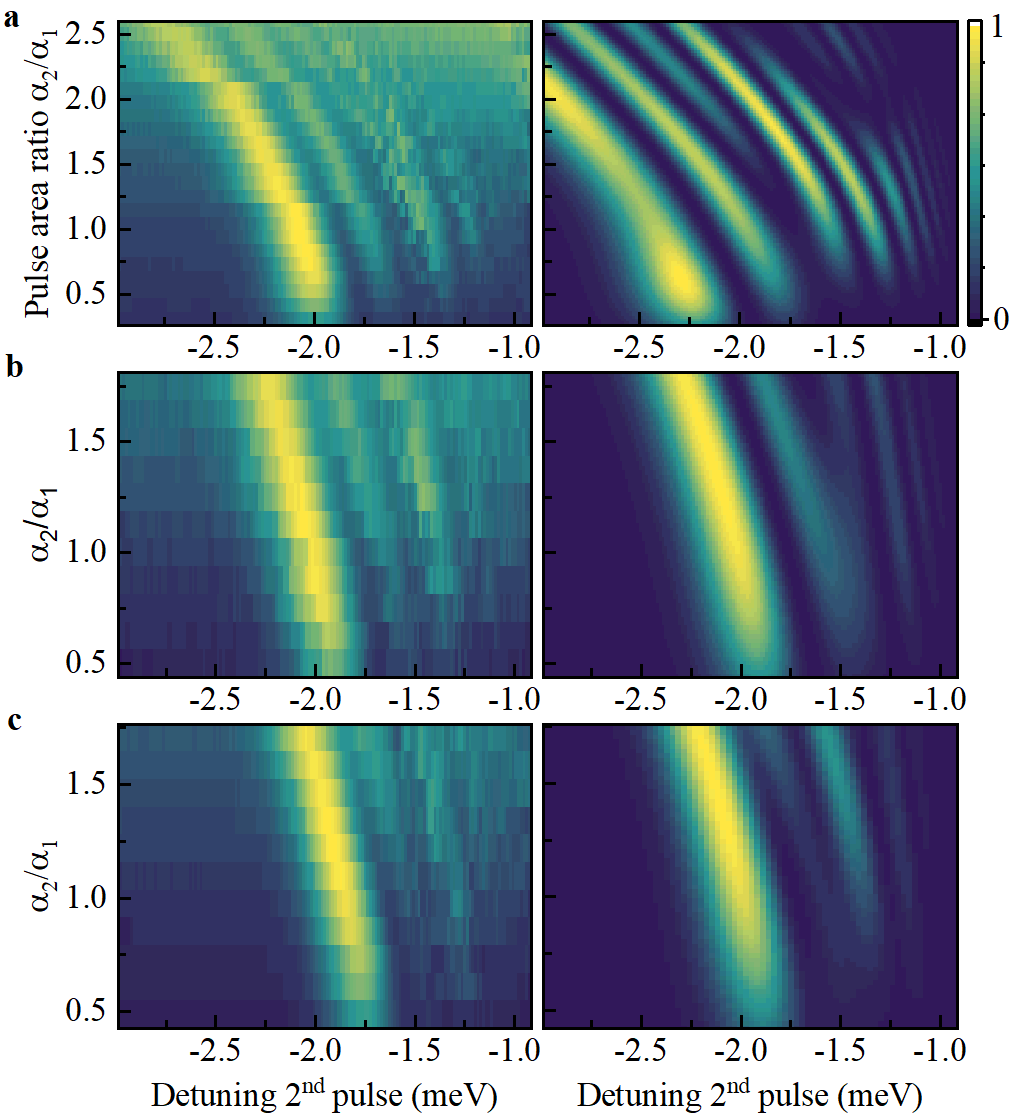}
    \caption{Trion emission intensity plotted as a two-dimensional colormap in dependence of detuning and intensity of the second pulse given in terms of pulse area ratio $\alpha_2 / \alpha_1$, each with fixed parameters for the first laser and for pulse length of $\SI{10}{\pico \second}$. Both experimental data (left) and simulations (right) are shown. First laser at (a) $\hbar \Delta_1= \SI{-0.7}{\milli \electronvolt}$ and $\alpha_1= 11 \pi$, (b) $\hbar \Delta_1= \SI{-0.7}{\milli \electronvolt}$ and $\alpha_1 = 8 \pi$ and (c) $\hbar \Delta_1= \SI{-0.7}{\milli \electronvolt}$ and $\alpha_1 = 8 \pi$ with a $\SI{4}{\pico \second}$ time delay for the second laser. Resonances can be observed with clear maxima revealing population inversions up to $0.67$.}
    \label{fig:graph2}
\end{figure}

To identify for which combination of parameters the population inversion is successful, we monitor the emission intensity of the trion transition while changing different pulse parameters of the swing-up excitation. As the parameter space is multi-dimensional, we record 2D colormaps as presented in Fig.~\ref{fig:graph2} where we scan the detuning and the laser intensity of the second laser pulse to find resonances and characteristics of the system, while we vary another third parameter from map to map. The integrated trion emission is normalized to the maximal intensity of the specific parameter set for better visibility in each map. As a first step we keep the pulse lengths at $\SI{10}{\pico \second}$ and fix the detuning and intensity of the first pulse to $\hbar \Delta_1 = \SI{-0.7}{\milli \electronvolt}$ and $\alpha_1 = 8 \pi$, respectively. The intensity of the pulse is normalized to the power of a resonant pulse with an area of $1\pi$, determined from resonant Rabi oscillations. We scan the detuning and the intensity of the second laser pulse in the range where high population inversion is to be expected~\cite{Bracht2021}; from $\hbar \Delta_2 = \SI{-3}{\milli \electronvolt}$ to $\SI{-0.92}{\milli \electronvolt}$ and from $\alpha_2/\alpha_1 = 0.44$ to $1.81$. Results can be seen in Fig.~\ref{fig:graph2}b with both experimental data (left) and simulations (right) in good agreement. One main region of high excitation fidelity can be seen at around $\hbar \Delta_2 = \SI{-2}{\milli \electronvolt}$, slightly shifting towards larger detunings with higher intensity. Within, we find a maximal population inversion at $\alpha_2 /\alpha_1 =1.1$ and $\hbar \Delta_2 =\SI{-2.05}{\milli \electronvolt}$ and estimate the fidelity to $0.66$ as the ratio of swing-up intensity to emission intensity under excitation with a resonant $\pi$-pulse. Notably, simulations suggest a maximum attainable efficiency of $0.97$. Towards lower detunings, additional weaker resonances can be seen. These clear oscillations in intensity over a wide detuning range indicate the coherent nature of the swing-up excitation and confirm that the emission does not originate from one of the laser pulses alone. 
In order to see the impact of the intensity of the first laser pulse, we now increase $\alpha_1$ to $11 \pi$ (Fig.~\ref{fig:graph2}a). While the signatures are similar to the lower-intensity case, their structure is more complex and features more resonances. In the experiment, we reach a similar efficiency of $0.67$ with $\alpha_2 /\alpha_1 = 0.93$  and $\hbar \Delta_2 =\SI{-2.08}{\milli \electronvolt}$. However, a minor mismatch between simulation and experiment can be observed: in the simulation, the regions of high intensity are shifted to larger negative detunings of the second laser with respect to the measured data. This shift increases as the pulse area of the second pulse is increased. In addition, multiple maxima appear in the simulated data set which are absent in the measured data. This small deviation between simulations and experiment is likely to result from a wavelength-dependent coupling of the detuned lasers to the weak cavity of the sample, forming between the distributed Bragg reflector beneath the QD and the sample surface. Since the power of the detuned lasers is calibrated to the pulse area needed for a resonant Rabi oscillation, a wavelength-dependent coupling of the laser leads to a small error in the pulse area, which due to the non-linear shift of the swing-up resonances results in small differences in the colormaps.
For high pulse areas in a region where $\alpha_2 /\alpha_1 \in [2, 2.5]$ and small detuning $\hbar \Delta_2 < \SI{-5}{\milli \eV}$ we observe a background, i.e. emission of the $X^-$ transition independent of the detuning of the second laser. The background implies non-coherent excitation induced by the high laser intensities and thus, for investigating the quality of the emitted photons using lower laser intensities is beneficial. Note that on the other hand too low pulse intensity decreases the efficiency of the scheme, making it challenging to find an optimal parameter set for the studied sample~\cite{Supplemental}. 
To further explore the multi-dimensional swing-up parameter space, we now turn to the effect of the time delay between the two pulses, for which we delay the second pulse by $\SI{4}{\pico \second}$ with respect to the first one (Fig.~\ref{fig:graph2}c). Consistent with the simulations, we see that the general characteristics of the swing-up excitation do not change much for an additional delay of $\SI{4}{\pico \second}$, i.e. we find the highest population efficiency to be $0.61$ at $\alpha_2 / \alpha_1 = 1.33$ and $\hbar \Delta_2 = \SI{-1.94}{\milli \electronvolt}$ which is similar to the maximum without delay. This is beneficial for applications that use this technique as it proves robust to fluctuations in the timing of the two pulses. Simulations show that the excitation works efficiently as long as the two pulses have a significant overlap~\cite{Supplemental} while being symmetric with respect to positive and negative delay~\cite{Bracht2021}. It is worthwhile to note that Bracht \textit{et al.} suggest that with zero time delay a maximal fidelity of $0.90$ can be achieved and a pulse separation of $\SI{2.5}{\pico \second}$ would achieve the highest population inversion for their studied pulse parameter set~\cite{Bracht2021}, while simulations with our experimental parameter set show that near-unity population of $0.97$ could be achieved with zero delay. 

\begin{figure}
    \centering
    \includegraphics[width=\linewidth]{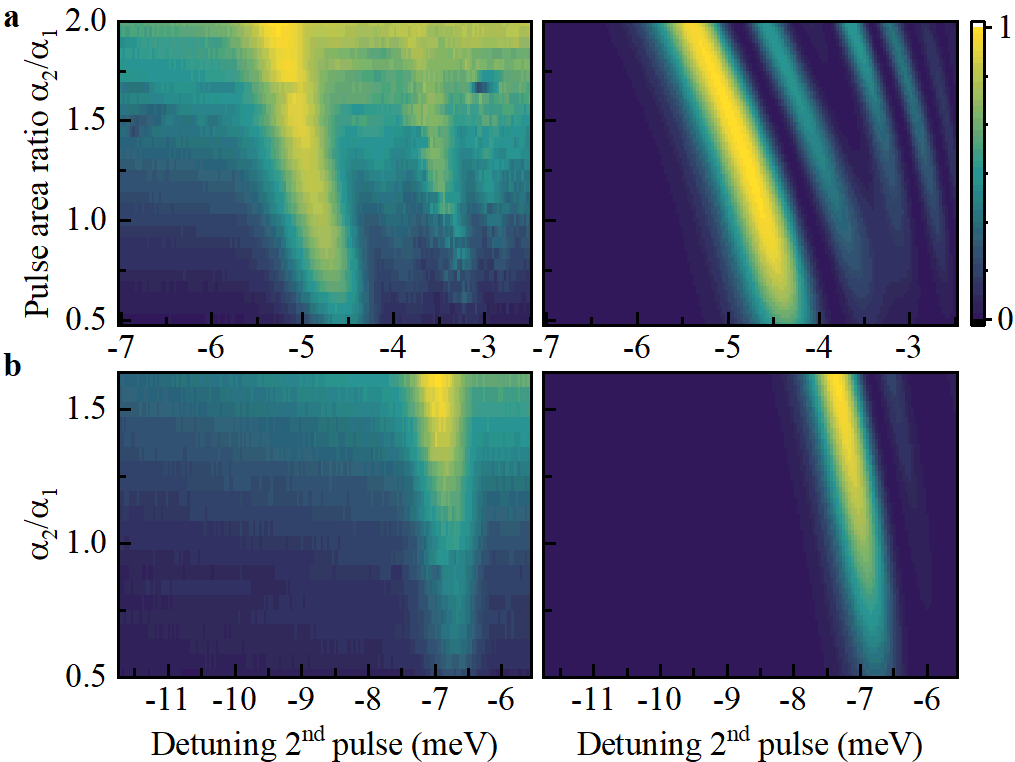}
    \caption{Experimental (left) and simulated (right) swing-up excitation fidelity dependent on detuning and intensity of the second pulse for fixed parameters of the first pulse at $\SI{5}{\pico \second}$ pulse duration. First laser is centered at (a) $\hbar \Delta_1 = \SI{-1.65}{\milli \electronvolt} $ and $\alpha_1= 9 \pi$, (b) $\hbar \Delta_1 = \SI{-3}{\milli \electronvolt}$ and $\alpha_1= 9 \pi$. With increasing detuning, the areas of highest population inversion are shifted towards higher excitation intensity.}
    \label{fig:graph3}
\end{figure}

To continue the study of the phase space, we investigate shorter pulses, i.e. $\SI{5}{\pico \second}$ on a second QD. Experimental results and simulation with the first laser at $\hbar \Delta_1 = \SI{1.65}{\milli \electronvolt}$ and $\alpha_1 = 9 \pi$ can be seen in Fig.~\ref{fig:graph3}a and show good agreement and similar resonances as for longer pulses. Note that the power of the resonant $\pi$-pulse to which the intensities are normalized is higher by a factor of $3$ compared to the power for the $\SI{10}{\pico \second}$ pulses due to their decreased spectral overlap with the narrow QD transition. In agreement with the measurements performed with higher excitation intensities (Fig.~\ref{fig:graph2}), high laser power induces a discrepancy between the resonance lines of experiment and simulation. In addition, in the experiment the point of highest occupation appears to be shifted towards higher pulse intensities, i.e. to a point above $\alpha_2 / \alpha_1 = 2.0$ which is outside of the measured region and where the emission background between the resonance lines is non-negligible. Compared to excitation with $\SI{10}{\pico \second}$ pulses, we see that the maximum population inversion is strongly shifted towards higher detunings and higher intensities of the second laser which is experimentally challenging and requires further engineering of the sample design. Furthermore, there are higher constraints for shorter pulses as larger detunings are required to prevent spectral overlap with the fundamental transition and as the short pulse duration leads to a higher impact of the relative pulse delay. 
Increasing the detuning of the first laser pulse (Fig.~\ref{fig:graph3}b) while keeping all other parameters constant results in fewer resonances that shift to higher detuning of the second laser as predicted by theory~\cite{Bracht2021}. From simulations it is expected that the maximum occupation is shifted to higher powers, which can not be confirmed experimentally as the area of maximal occupation is shifted outside of the measured region.
Taken together, for the studied sample we find a trade-off between best efficiency and minimal background at a fidelity of $0.67$ with the first pulse set to $\hbar \Delta_1 = \SI{-0.7}{\milli \electronvolt}$ and $\alpha_1 = 11 \pi$, and the second pulse set to  $\alpha_2/\alpha_1 = 0.94$  and $\hbar \Delta_2 = \SI{-2.08}{\milli \electronvolt}$ for $\SI{10} {\pico \second}$ pulses.

Having found optimal parameters for the excitation technique and sample under investigation we now proceed to characterize the quality of the single photons emitted from the QD under swing-up excitation. To benchmark the characteristics, we compare the results with emission under the established coherent resonant excitation.

\begin{figure}
    \centering
    \includegraphics[width=\linewidth]{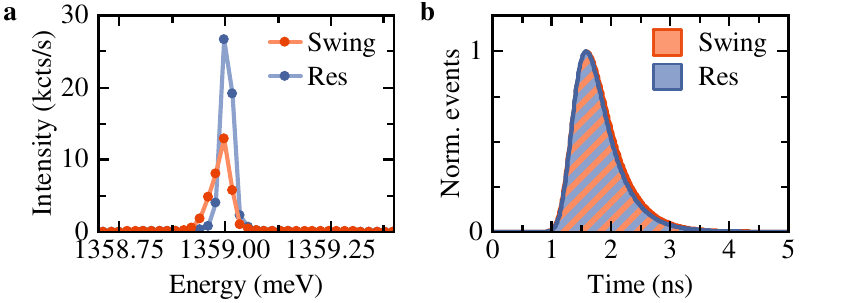}
    \caption{Emission spectra (a) and radiative lifetime (b) of the trion transition under swing-up excitation (orange) compared to resonant excitation (blue). The emission peak under swing-up excitation is slightly broadened and shifted due to effects of the high laser intensities on the surrounding solid-state environment. The radiative lifetimes of emission under both excitation methods are comparable.}
    \label{fig:graph4}
\end{figure}

To gain initial insight into the quality of the swing-up photons, we take a look at the emission spectrum taken with a spectrometer (Fig.~\ref{fig:graph4}a, orange line). Compared to the emission peak under resonant excitation (Fig.~\ref{fig:graph4}a, blue line) the integrated intensity is reduced to $0.67$, showing the lower efficiency of the excitation method. However, note that the measurements were taken in a setup where the emitted light in the detection path is filtered by a linear polarizer, reducing the intensity by $0.5$ overall. This is needed for the measurements with resonant excitation to separate QD emission and excitation laser while it would not be necessary for the excitation via swing-up due to spectral detuning of excitation and emission. Two more observations can be made comparing the emission to the one under resonant excitation: The spectrum is shifted by about $\SI{11}{\micro \electronvolt}$ and slightly broadened. We attribute this to a susceptibility of the semiconductor environment to the detuned but strong laser pulses required for swing-up excitation. To verify this, we analyze the $X^-$ emission under phonon-assisted excitation and the $X^0$ emission of the biexciton cascade under two-photon excitation with increasing excitation power. In both cases we observe similar characteristics (Supplemental Material ~\cite{Supplemental}). The shift can be explained by re-normalization of the resonance due to localized heating of the material, while the broadening can be attributed to spectral diffusion induced by the strong laser pulses. Nevertheless, as both artifacts increase with laser power, we aim to stay at low excitation powers to ensure a high quality of the emitted photons. Time-resolved photoluminescence measurements of both swing-up and resonant excitation are presented in Fig.~\ref{fig:graph4}b and confirm comparable values of the lifetimes of $450 \pm \SI{14}{\pico \second}$ and $419 \pm \SI{13}{\pico \second}$, respectively.

\begin{figure}
    \centering
    \includegraphics[width=\linewidth]{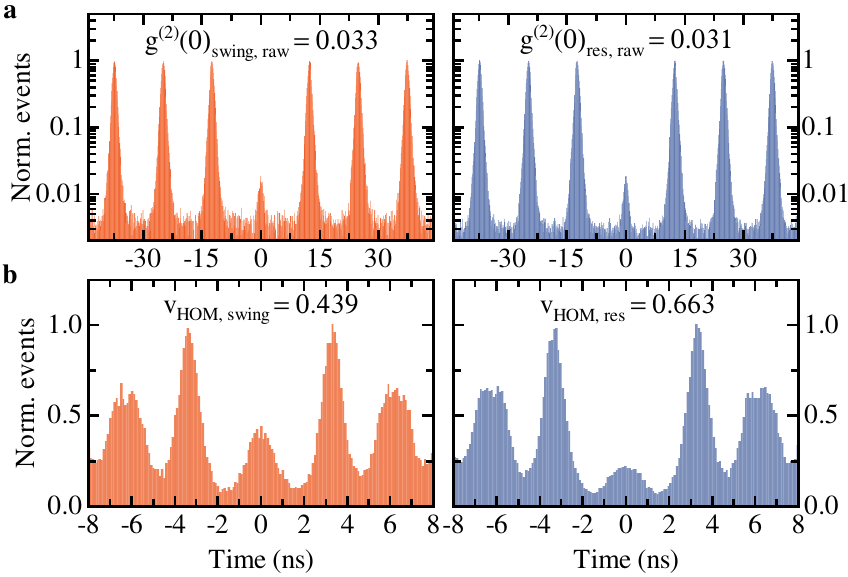}
    \caption{Single-photon purity and indistinguishability of single photons emitted under swing-up excitation. (a) Second-order correlation function/histogram of swing-up excitation (orange) and resonant excitation (blue). Both raw values of $g^{(2)}(0)_{\text{swing,raw}} = 0.033 \pm 0.001$ and $g^{(2)}(0)_{\text{res,raw}} = 0.031\pm 0.001$ are in excellent agreement. (b) Correlation measurements of the swing-up excitation (orange) show a visibility of $v_{\text{HOM,swing}} = 0.439^{+0.047}_{-0.049}$ which is lower than the value of $0.663^{+0.032}_{-0.035}$ obtained for resonant excitation (blue). This is attributed to the spectral broadening of the emission observed in Fig.~\ref{fig:graph4}.}
    \label{fig:graph5}
\end{figure}

To study the single-photon purity we measure the second-order correlation function $g^{(2)}( \tau)$. The time-resolved correlations are presented in Fig.~\ref{fig:graph5}a, yielding a raw value of $g^{(2)}(0)_{\text{swing,raw}}= 0.033 \pm 0.001$ for swing-up excitation (orange). Taking the background of the SPADs into account (Supplemental Material~\cite{Supplemental}), we achieve a corrected value of $g^{(2)}(0)_{\text{swing, corr}} =0.013 \pm 0.001$ which is so far the lowest $g^{(2)}(0)$ value for swing-up excitation, likely due to the optimized excitation conditions. This is in excellent agreement with the raw value of $g^{(2)}(0)_{\text{res,raw}} = 0.031\pm 0.001$ for resonant excitation (Fig.~\ref{fig:graph5}a right), implying that there are no limitations on the single-photon purity from the swing-up scheme itself. The limiting factor for the single-photon purity in this case is the re-excitation which is especially relevant for excitation pulses which are long with respect to the radiative lifetime \cite{Fischer2018a}. 

We conclude our investigation of the quality of the emitted single photons with the study of their indistinguishability. To this end, we measure correlations of consecutively emitted photons in an unbalanced Mach-Zehnder interferometer. The correlation histogram for photons under swing-up excitation is depicted in Fig.~\ref{fig:graph5}b on the left. Evaluating the Hong-Ou-Mandel-visibility $v_{\text{HOM}}$ by comparing the center peak to the neighbouring peaks (Supplemental Material~\cite{Supplemental}) we obtain a value of $v_{\text{HOM,swing}} = 0.439^{+0.047}_{-0.049}$ which, to the best of our knowledge, is the first measured indistinguishability for photons emitted by a QD under swing-up excitation. This is lower than the value of $v_{\text{HOM,res}} = 0.663^{+0.032}_{-0.035}$ obtained for resonant excitation. We attribute this degradation under swing-up excitation to the broadened emission peak and thus decreased spectral overlap caused by the high laser intensities.
We note that for our sample and resonant excitation, the indistinguishability of trion emission is reduced compared to the neutral exciton emission which yields $v_{\text{HOM,res,}X^0} = 0.821^{+0.019}_{-0.018}$ ~\cite{Supplemental}, indicating the presence of co-tunneling between the QD and the n-doped layer. However, using swing-up with uncharged QDs adds further complexity due to the overlap of swing-up resonances of the neutral exciton and two-photon swing-up resonances of the biexciton. Consequently, to exploit the full potential of the swing-up excitation further optimized samples are required which exhibit less co-tunneling and less spectral diffusion resulting from detuned but strong laser pulses.

In summary, we have investigated the recent swing-up excitation method~\cite{Bracht2021} in detail by studying the multi-dimensional parameter space spanned by the two laser pulse parameter sets with respect to feasibility in the experiment and quality of the emitted single photons. The scheme is very promising as it uses two far red-detuned laser pulses for a coherent population transfer due to beating of the frequencies with predicted near-unity fidelity. In general, the excitation method works well with up to $0.67$ efficiency compared to a resonant $1\pi$ pulse within a certain parameter range that is currently fixed by experimental limitations. We find a near-perfect single-photon purity with $g^{(2)}(0)_{\text{swing,corr}} =0.013 \pm 0.001 $. In addition, we measured the indistinguishability yielding a value of $v_{\text{HOM,swing}} = 0.439^{+0.047}_{-0.049}$ which is limited by the studied sample, in particular by co-tunnelling and spectral diffusion caused by the high laser powers. However, with optimized samples we expect the scheme to play a major role in the generation of complex non-classical states of light, such as photonic graph states, which require a combination of coherent excitation, high brightness, high indistinguishability and absence of polarization filtering.

We gratefully acknowledge financial support from the German Federal Ministry of Education and Research via the funding program Photonics Research Germany (Contract No. 13N14846), the European Union's Horizon 2020 research and innovation program under Grants Agreement No. 862035 (QLUSTER) and No. 899814 (Qurope), the Deutsche Forschungsgemeinschaft (DFG, German Research Foundation) via the projects MQCL (INST 95/1220-1), CNLG (MU 4215/4-1), TRR 142 (grant No. 231447078), and Germany's Excellence Strategy (MCQST, EXC-2111, 390814868), the TUM Institute for Advanced Study, the Bavarian State Ministry of Science and Arts via the project EQAP and the Exploring Quantum Matter (ExQM) program funded by the state of Bavaria.

\newpage

\onecolumngrid

\section{Supplemental Material}
\section{Theoretical model}

In our simulations, a two level system consisting of a ground state $\ket{0}$ and an excited state $\ket{1}$ with an energy separation between the two states of $\hbar\omega_0$ and a bichromatic driving field $E(t)$ are considered within the rotating wave approximation. Assuming that the field is linearly polarized in x-direction, we define a time-dependent term $\Omega(t) = - d_x E(t)/\hbar$, where $d_x$ is the transition dipole moment in x-direction. The total excitation field consisting of two independent Gaussian pulses is given by 
\begin{equation}
\Omega(t) = \Omega_1(t)e^{-i\omega_1 t} + \Omega_2(t+\tau)e^{-i\omega_2 t+i\phi}    
\end{equation}
where $\omega_{1,2}$ are the center frequencies of the two pulses, $\tau$ and $\phi$ are time delay and phase difference between the two pulses, respectively. Note that the phase does not affect the scheme and is neglected for further calculations ~\cite{Bracht2021}. Real Gaussian pulse envelopes $\Omega_{1,2} (t)$ are given by $\alpha_{1,2} / \sqrt{2\pi\sigma_{1,2}^2} \exp{-t^2/(2\sigma_{1,2}^2)}$, where $\sigma_{1,2} = \text{FWHM}_{1,2} / \sqrt{4 \ln{2}}$ are described by the intensity full width half maximum ($\text{FWHM}_{1,2}$) of each pulse, and $\alpha_{1,2}$ are pulse areas defined by $\alpha_{1,2} = \int_{-\infty}^{+\infty} \Omega_{1,2}(t) \,dt$. Thus we can describe the system Hamiltonian by
\begin{equation}
    H
    = \hbar\omega_0 \sigma^\dagger\sigma + \frac{\hbar}{2}\Omega(t)\left(\sigma
    + \sigma^\dagger \right),
\end{equation}
where the annihilation operator of the two level system is defined by $\sigma = \ketbra{0}{1}$. For further calculations, we apply the rotating wave approximation and move to the frame rotating at the center frequency of the closer detuned laser pulse. To study the dynamics and the final population in the excited state, we solve the Von-Neumann equation
\begin{equation}
\frac{\partial}{\partial t} \rho (t)= - \frac{i}{\hbar}[H,\ \rho(t)]
\end{equation}
within the time range from $t_{i}$ to $t_{f}$. The time-dependent density operator $\rho(t)$ is initialized in the ground state, i.e., $\rho(t_{i}) = \ketbra{0}$. We calculate the final excited state population as $\bra{1}\rho(t_{f})\ket{1}$. For the calculations, we use a finite time window of $\SI{80}{\pico \second}$ exceeding the time of the population inversion.

\section{Sample and experimental setup}

The sample used for our measurements consists of self-assembled InGaAs QDs grown by molecular beam epitaxy. The layer of dots are enclosed by a diode to further control and stabilize the environment~\cite{warburton2000, seidl2005}, while a distributed Bragg reflector beneath the QDs enhances the collection efficiency.
The sample is placed inside an exchange-gas dip-stick located in a liquid-helium bath resulting in a constant temperature of $\SI{4.2}{\kelvin}$. Optical access is granted from the top using a state-of-the art cross-polarized resonance fluorescence confocal microscope~\cite{Kuhlmann2013}. High-precision positioning of the QD with respect to the laser field is achieved via a stack of nanopositioners. For excitation, we use a $\SI{150}{\femto \second}$ Ti:Sapph laser whose output is split into two pulses and shaped using two fully independent $4 f$ pulse-shapers~\cite{Weiner1988}. Full control of the sub-picosecond time delay between the two pulses is achieved by a delay line. Simultaneously, two separate optical attenuators allow independent control of the pulse area of both pulses after which the pulses are recombined and sent to the confocal microscope. This experimental setup enables us to individually control frequency, pulse duration and intensity of the two pulses and their relative time delay. The emitted photons are frequency filtered and detected by either a spectrometer and a CCD camera or by single-photon avalanche diodes (SPADs).

\section{Additional simulations}
\subsection{Simulations for different pulse areas of the first pulse}

\begin{figure}
    \centering
    \includegraphics[width=\linewidth]{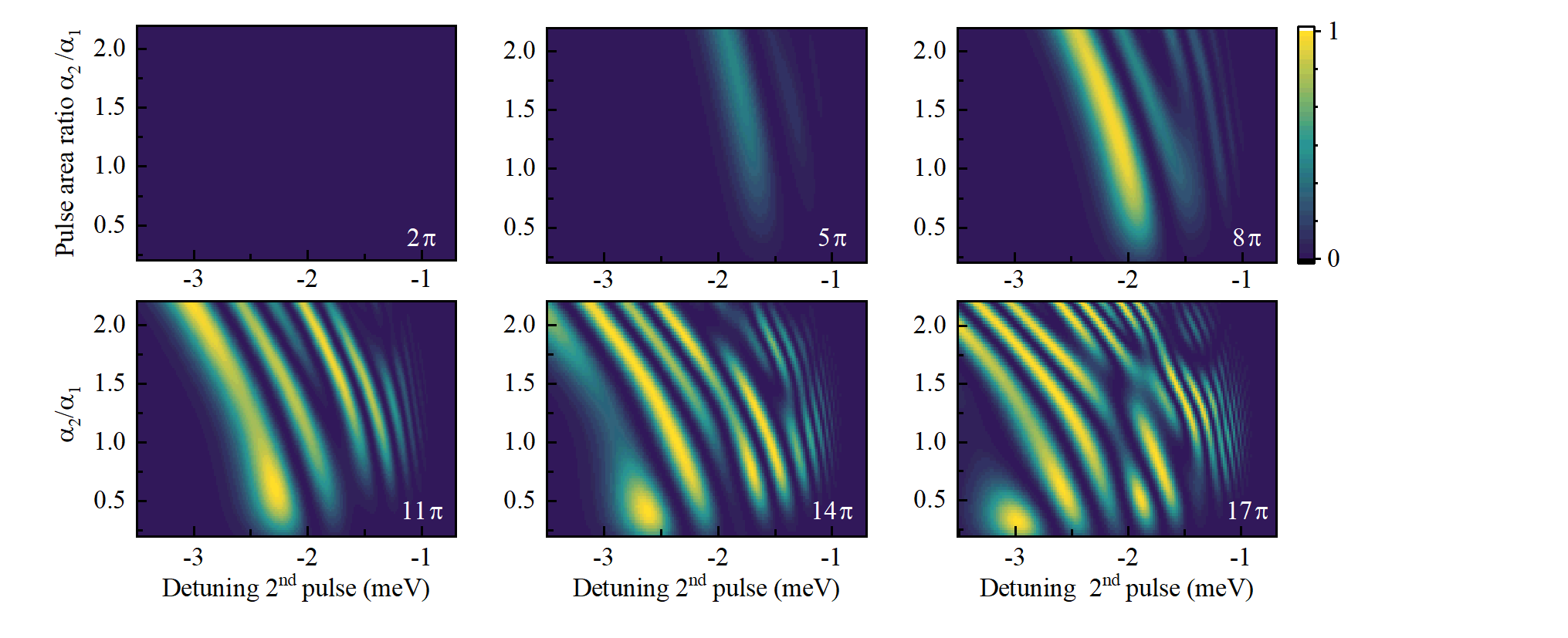}
    \caption{Simulation series for increasing pulse area of the first pulse from $\alpha_1=2 \pi$ up to $\alpha_1=17 \pi$. Towards lower pulse intensity the total efficiency of the scheme decreases, while multiple and more complex resonances appear with larger pulse areas. The detuning and the duration of the first pulse are fixed to $\Delta_1 = \SI{-0.7}{\milli\eV}$ and $\SI{10}{\pico\second}$, respectively, while they are swept for the second laser pulse.}
    \label{fig:supp_graph1}
\end{figure}

In Fig.~\ref{fig:supp_graph1} we present additional simulations we performed for varying pulse areas from $\alpha_1= 2\pi$ to $17\pi$ of the first laser pulse to provide a more detailed analysis of the multidimensional parameter space. 
We keep the other parameters in the simulation fixed to the values also used in the main manuscript, i.e. $\SI{10}{\pico\second}$ pulse duration, $\Delta_1=\SI{-0.7}{\milli\eV}$ detuning of the close detuned laser and $\SI{0}{\pico\second}$ time delay. Equivalent to the main manuscript, we perform two-dimensional parameter sweeps of the second laser changing relative pulse area and detuning to determine optimal resonance conditions for the different first laser pulse areas.
As stated in the main manuscript, we observe for lower pulse area of the first laser $\alpha_1<8 \pi$ the preparation fidelity of the swing-up scheme decreases drastically, while near-unity inversion can be achieved for pulse areas of the first laser pulse $\alpha_1 \geq 8 \pi$. In addition, for very large pulse areas of both laser pulses we observe an increased complexity in resonance lines (lower part of Fig.~\ref{fig:supp_graph1}), featuring multiple areas of high-fidelity population inversion for different detunings and pulse areas of the second laser pulse.

\subsection{Simulations for varying time-delay between pulses}

\begin{figure}
    \centering
    \includegraphics[width=\linewidth]{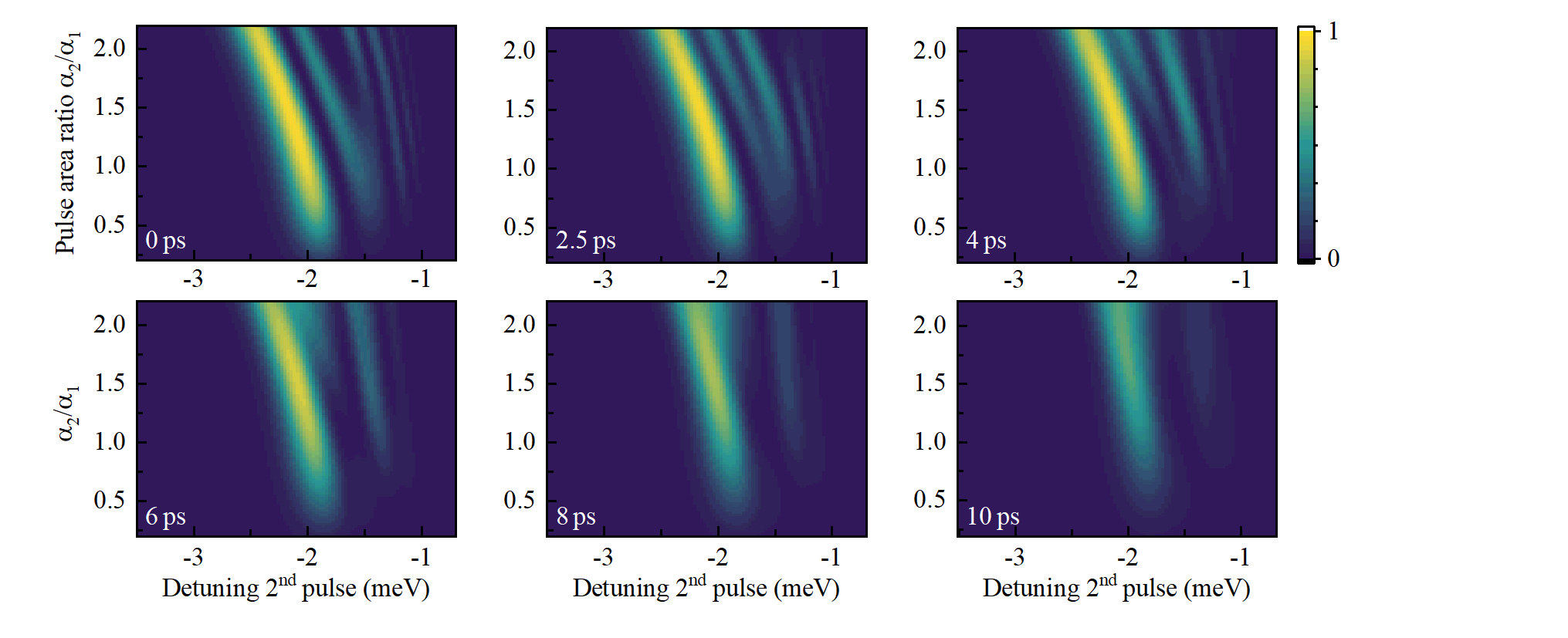}
    \caption{Simulation series investigating the influence of the time delay between the pulses. The relative delay of the second pulse is tuned between the figures from $\SI{0}{\pico\second}$ to $\SI{10}{\pico\second}$. The detuning and pulse area of the first laser pulse are fixed at $\Delta_1= \SI{-0.7}{\milli\eV}$ and $\alpha_1= 8 \pi$. With increasing time delay, the maximum population inversion fidelity decreases, while the position of the maximal efficiency does not change significantly with respect to detuning and pulse power of the second laser. }
    \label{fig:supp_graph2}
\end{figure}

In the main part of the manuscript we state that the population inversion fidelity is not highly sensitive on the time delay between the two pulses, provided they have a significant temporal overlap. In Fig.~\ref{fig:supp_graph2} we present simulations with increasing relative time delay of the second pulse from $\SI{0}{\pico\second}$ to $\SI{10}{\pico\second}$ where the pulse durations of both laser pulses are fixed at $\SI{10}{\pico\second}$ and the parameters of the first laser are given by $\Delta_1 = \SI{-0.7}{\milli\eV}$ and $\alpha_1= 8 \pi$. While from $\SI{0}{\pico\second}$ up to $\SI{4}{\pico\second}$ time delay the population can be successfully inverted with near-unity fidelity, the efficiency of the population transfer decreases drastically for larger time delays, as the overlap of the two pulses, needed for the swing-up effect, decreases. The position of maximal efficiency of the scheme with respect to the detuning and pulse area of the second pulse does not change significantly with increasing time delay. Note that a short relative time delay of $\SI{2.5}{\pico\second}$ between the pulses does not improve the efficiency for our chosen parameter set, highest population inversion is reached for no delay.

\subsection{Simulations for large detunings of first pulse}

\begin{figure}
    \centering
    \includegraphics[width=\linewidth]{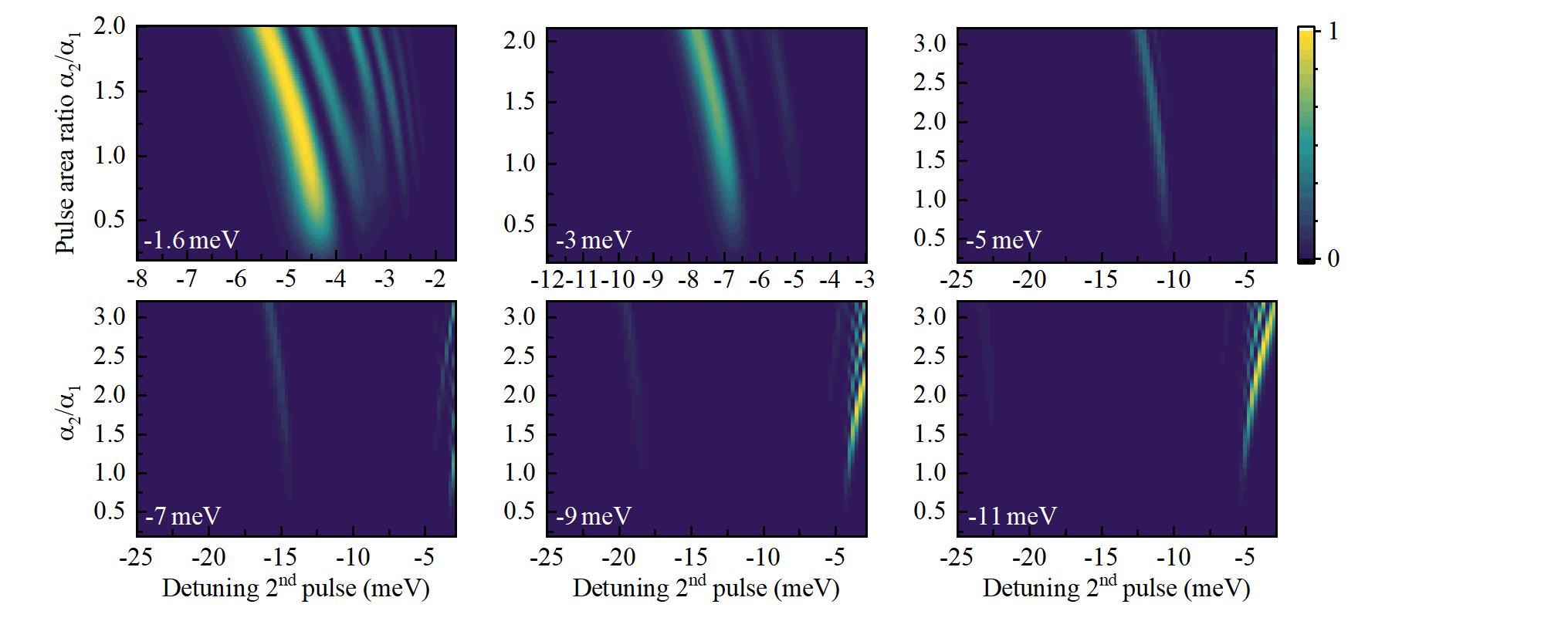}
    \caption{Simulation series for increasing detuning of the first pulse, from $\Delta_1= \SI{-1.6}{\milli\eV}$ up to $\SI{-11}{\milli\eV}$. Larger detuning of the first laser pulse shifts the area of maximum inversion fidelity to larger pulse area and higher detuning of the second laser pulse. The for small detunings of the second pulse that appear for $\Delta_2= \SI{-9}{\milli\eV}$ and $\SI{-11}{\milli\eV}$ correspond to typical swing up resonances when the second pulse is less detuned from the two-level system than the first one.}
    \label{fig:supp_graph3}
\end{figure}

We present simulations with increasing detuning of the first laser pulse from $\Delta_1= \SI{-1.6}{\milli\eV}$ up to $\SI{-11}{\milli\eV}$ (Fig.~\ref{fig:supp_graph3}). The pulse area of the first laser pulse is fixed at $\alpha_1= 9 \pi$, while the duration is set to $\SI{5}{\pico\second}$ for both pulses and a relative time delay of zero is considered. As stated in the main manuscript, we observe that larger detunings decrease the efficiency of the scheme and the point of high efficiency shifts to larger detunings and pulse area of the second laser pulses.
The prominent additional lines observable for larger detuning of the first laser and small detuning of the second laser are the typical swing-up resonances, appearing as the detuning of the second pulse here is smaller than the detuning of the first pulse which leads to swapped roles of the two pulses.

\section{Emission peak broadening under two-photon excitation and phonon assisted excitation for high excitation powers}

\begin{figure}
    \centering
    \includegraphics[width=\linewidth]{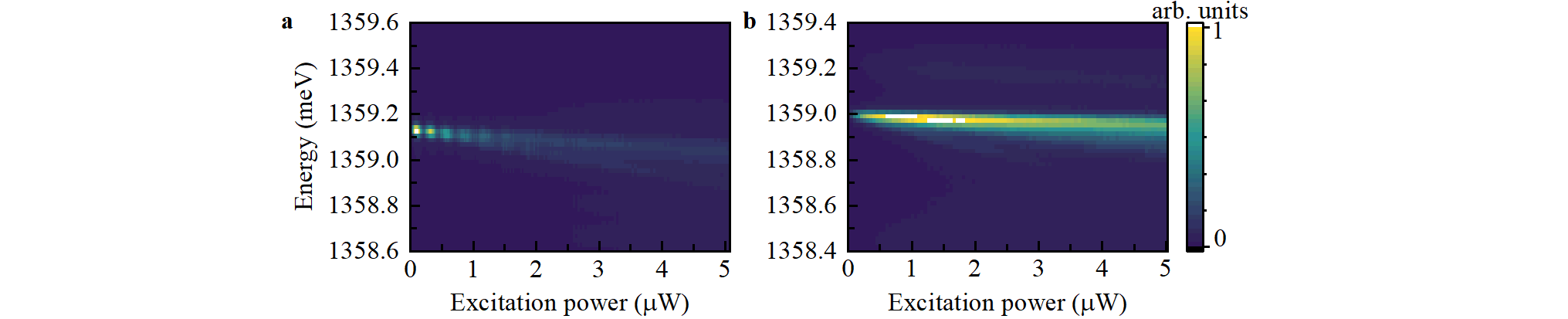}
    \caption{Emission intensity of the neutral exciton $X^0$ under TPE via the biexciton (a) and emission of the negatively charged trion $X^-$ under PAE. In both measurements the photoluminescence emission of the quantum dot broadens and shifts towards lower energy.}
    \label{fig:supp_graph5}
\end{figure}

We state that the spectral broadening of the trion emission under swing-up excitation is an effect of the high laser powers and not the excitation scheme itself. Additionally the observed broadening increases with increasing excitation power. To confirm that this effect is independent of the swing-up excitation scheme, we present additional measurements we performed of the emission of the trion and the neutral exciton under both phonon-assisted excitation (PAE) and two-photon excitation (TPE), respectively. The results are presented in Fig.~\ref{fig:supp_graph5}.

For TPE we observe  Rabi oscillations with increasing laser power of up to $\SI{1}{\micro\watt}$. They are accompanied by a minor spectral shift in emission. For larger laser powers of up to $\SI{5}{\micro\watt}$ the powerspectrum exhibits a clearly visible asymmetric broadening that scales linearly with the excitation power. For the excitation, the pulsed laser was centered exactly between the exciton and biexciton emission frequencies.

Similar broadening is observed in the emission of the trion under pulsed PAE. Here, the laser was centered at $\SI{1365.18}{\milli\eV}$, blue-detuned from the transition.

As we observe this effective broadening of the quantum dot emission under high excitation powers independent of the excitation scheme, we attribute this to the susceptibility of the semiconductor environment to the strong laser pulses.

\section{Correlation histogram analysis}

\begin{figure}
    \centering
    \includegraphics[width=\linewidth]{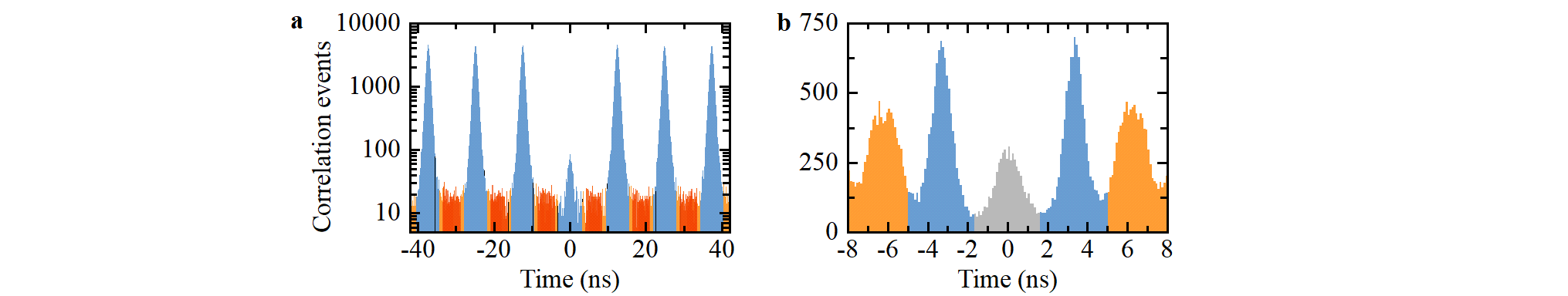}
    \caption{Exemplary (a) $g^{(2)}(t)$ and (b) HOM correlation histograms of emission under swing-up excitation.}
    \label{fig:supp_graph6}
\end{figure}
For further analysis of the single-photon purity, the raw time-tagged correlation data are integrated into bins of $\SI{100}{\pico\second}$ as exemplarily shown for emission under swing-up excitation in Fig.~\ref{fig:supp_graph6}a. To obtain the value of the raw $g^{(2)}(0)$, we integrate the counts of the center peak as well as six neighbouring peaks within a time window of $\SI{6}{\nano\second}$ each, as marked in blue. This way, the integration regions are kept to a minimum while ensuring that the whole peaks are included into the analysis for the most exact result. The ratio of the integrated intensity of the center peak and the average of the side peak gives $g^{(2)}(0)_{\text{raw}}$.
For the corrected background-subtracted values the six background regions of $\SI{4.5}{\nano\second}$ between the peaks (depicted in red) are averaged and normalized with respect to the integration window, and subtracted from the data. 
Assuming Poissonian noise statistics for the detection events, the error for the raw values are given by the standard error $\text{error}_{\text{raw}} = \text{A}_{\text{center,raw}}^{-1/2}* {g^{(2)}}_{\text{raw}}$ where $\text{A}_{\text{center,raw}}$ is the integrated area of the central peak. For the error of the background-corrected values we use the background-included center peak area as the background accounts for a non-trivial portion of the center peak, leading to $\text{error}_{\text{corr}} = \frac{\text{A}_{\text{center,raw}}}{\text{A}_{\text{avg.neighbour,raw}}^2}$.

For the indistinguishability measurements we adjust the pulse separation to $\SI{3.3}{\nano \second}$ to minimize the impact of charge noise on the result~\cite{Thoma2016}. Again, we integrate the raw correlation data into time-bins of $\SI{100}{\pico\second}$ as shown in Fig~\ref{fig:supp_graph6}b for the exemplary case of emission of the trion under swing-up excitation. The HOM-visibility is given by $v_{\text{HOM}}=1-\frac{\text{2A}_{\text{center}}}{\text{A}_{\text{left}}+\text{A}_{\text{right}}}$ where the center peak (shown in grey) and its two neighbouring peaks (blue) are integrated each within a window of size that equals half the distance between the two neighbouring peaks. We estimate the error by Poissonian statistics and by taking into account the timing-jitter of the detectors of $\SI{300}{\pico\second}$. This we do by adding and subtracting $\pm \SI{150}{\pico\second}$ to the integration window of the peaks.

\section{HOM-visibility measurements of the neutral exciton}

\begin{figure}
    \centering
    \includegraphics[width=\linewidth]{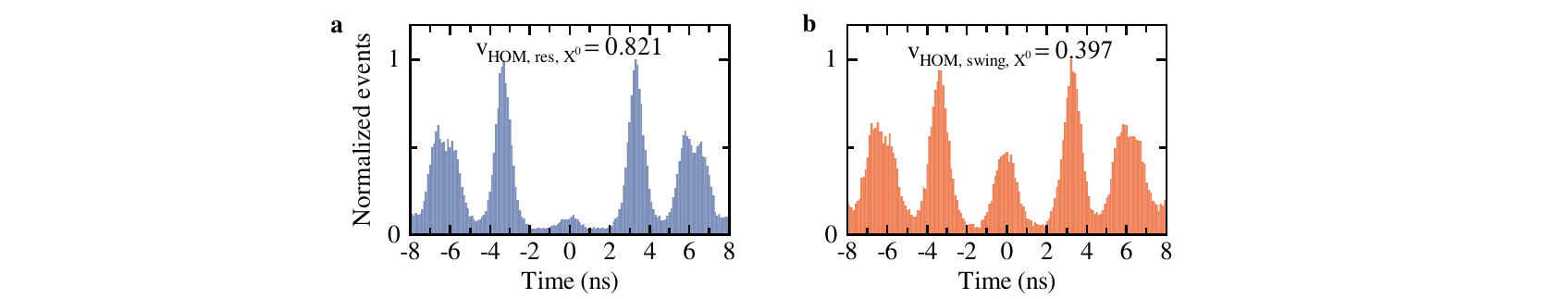}
    \caption{Indistinguishability measurements of the neutral exciton $X^0$ under (a) resonant and (b) swing-up excitation. The measured HOM-visibilities are $v_{\text{HOM,res,}X^0}=0.821^{+0.019}_{-0.018}$ and $v_{\text{HOM,swing,}X^0}=0.397^{+0.041}_{-0.041}$, respectively.}
    \label{fig:supp_graph4}
\end{figure}

HOM-visibility measurements on the negatively charged trion under both resonant excitation and swing-up excitation are shown in the main manuscript. 
To demonstrate that the indistinguishability of the trion is limited by the studied sample structure, i.e. originating from co-tunneling from the contacts of the diode, here we present HOM visibility measurements of the neutral exciton we performed on the same quantum dot (Fig.~\ref{fig:supp_graph4}). With the measured $v_{\text{HOM,res,}X^0}=0.821^{+0.019}_{-0.018}$ we achieve a much better indistinguishability under resonant excitation (Fig.~\ref{fig:supp_graph4}a) than for the negatively charged trion, indicating that the limited indistinguishability we observe originates from electron co-tunneling between the n-doped back contact and the quantum dot when setting the one-electron ground state. Interestingly, indistinguishability measurements we performed under the swing-up excitation scheme on the neutral exciton exhibit a prominent degradation of the visibility $v_{\text{HOM,swing,}X^0}=0.397^{+0.041}_{-0.041}$ (Fig.~\ref{fig:supp_graph4}b) compared to the resonant excitation case. We attribute this aggravation to two major causes: firstly, the strong electric fields of the swing-up excitation pulses disturb the local charge environment gravely, leading to a degradation of indistinguishability as similarly observed for the negative trion in the main manuscript. Secondly, we observe that with swing-up excitation on the neutral exciton, at areas of maximum population inversion of the exciton, the two swinger pulses additionally have a non-zero probability of exciting the biexciton. The subsequent cascaded emission leads to timing-jitter for the exciton emission, resulting in an additional deterioration of the HOM visibility~\cite{Scholl2020}. Thus, we find the swing-up excitation of the exciton state sub-optimal for benchmarking the photon emission properties. 

\providecommand{\noopsort}[1]{}\providecommand{\singleletter}[1]{#1}%

\newpage
\end{document}